\documentclass{ws-p10x7}
\newcommand{\be}{\begin{equation}}
\newcommand{\ee}{\end{equation}}
\newcommand{\ba}{\begin{eqnarray}}
\newcommand{\ea}{\end{eqnarray}}
\newcommand{\dis}{\displaystyle}

\newcommand{\re}{\mbox{Re}\,}
\newcommand{\im}{\mbox{Im}\,}

\begin{document}

\title{$\varepsilon_K'/\varepsilon_K$ at Next-to-Leading 
in $1/N_c$ and to Lowest Order CHPT}

\author{Johan Bijnens}

\address{Department of Theoretical Physics 2, Lund
University\\ S\"olvegatan 14A, S 22362 Lund, Sweden}

\author{Joaquim Prades}

\address{
Departamento de F\'{\i}sica Te\'orica
y del Cosmos, Universidad de Granada \\ Campus de Fuente Nueva,
E-18002 Granada, Spain}  

\twocolumn[\maketitle\abstract{We report on a calculation
of $\varepsilon_K'/\varepsilon_K$ at next-to-leading
order in the $1/N_c$ expansion and to lowest order
in Chiral Perturbation Theory. We also discuss the
chiral corrections to our results and give the
result of including the two known chiral corrections.}]

\section{Introduction}%1

Recently, 
direct CP violation in the Kaon system has been
unambiguously established by KTeV
at Fermilab and by NA48 at CERN \cite{KTEV}.
The present world average is 
\ba
{\re} \left(\varepsilon_K'/\varepsilon_K\right) \, 
&=& (19.3\pm2.4) \cdot 10^{-4}  \, .
\ea

Recent reviews and predictions for this quantity in the
Standard Model and earlier references
are in  \cite{lastresults}.
 Here, we report on a calculation   \cite{epsprime}
of this quantity in the chiral limit and next-to-leading (NLO) order
in  $1/N_c$. We also discuss the
changes when the  known chiral corrections 
-final state interactions
(FSI) and $\pi_0-\eta$ mixing- are included.

Direct CP-violation in the $K\to\pi\pi$ decay amplitudes
is parameterized by
\ba
\frac{\varepsilon_K'}{\varepsilon_K}&=&
\frac{1}{\sqrt 2}\left[ \frac{A\left[ K_L \to (\pi\pi)_{I=2}\right]}
{A\left[K_L \to (\pi\pi)_{I=0}\right]} 
\right. \nonumber \\
&-& \left. 
 \frac{A\left[ K_S \to (\pi\pi)_{I=2}\right]}
{A\left[K_S \to (\pi\pi)_{I=0}\right]} \right] \, .
\ea

$K\to\pi\pi$  amplitudes
can be decomposed into definite isospin amplitudes as
\ba
i \, A[K^0\to \pi^0\pi^0] &\equiv& {\frac{a_0}{\sqrt 3}} \, 
 e^{i\delta_0}
-\sqrt{\frac{ 2}{ 3}} \, 
a_2 \, e^{i\delta_2} \, , \nonumber \\
i \, A[K^0\to \pi^+\pi^-] &\equiv& {\frac{a_0}{\sqrt 3}} \, 
 e^{i\delta_0}
+\frac{a_2}{\sqrt 6} \, 
e^{i\delta_2}\, . 
\ea
with $\delta_0$ and $\delta_2$ the
FSI phases. 

We want to predict
$a_0$ and $a_2$ to NLO
order in $1/N_c$ and  to lowest order in CHPT.

\section{Short-Distance Scheme and Scale Dependence}

 The procedure to obtain the Standard Model effective action 
$\Gamma_{\Delta S=1}$   below the $W$-boson mass
has become standard
and  explicit calculations have been performed 
to two-loops\cite{two-loop}.
The full process implies choices of short-distance scheme,
regulators, and operator basis.
Of course, physical matrix elements cannot depend on these choices. 
 
The Standard Model $\Gamma_{\Delta S=1}$ effective action
at scales $\nu$ somewhat below the charm quark mass, takes
the form \cite{Buras}
\ba
\label{MSbar}
\Gamma_{\Delta S=1}&\sim& 
{\dis \sum_{i=1}^{10}}\, C_i(\nu)  \int {\rm d}^4 x \, Q_i(x)
+ {\rm h.c.} 
\ea
where  $Q_i(x)$ are four-quark operators
and $C_i = z_i + \tau \, y_i$ are Wilson coefficients.
In the presence of CP-violation, 
 $\tau \equiv - V_{td} V_{ts}^*/ V_{ud} V_{us}^*$
gets an imaginary part.

At low energies, it is more convenient to use an effective action
$\Gamma_{\Delta S=1}^{LD}$ which uses
different degrees of freedom. Different
regulators and/or  operator basis can be more practical too. 
The effective action  $\Gamma_{\Delta S=1}^{LD}$ 
depends on all these choices and in particular on the scale $\mu_c$
introduced to regulate the divergences,
analogous to $\nu$ in (\ref{MSbar}) and on
effective couplings
$g_i$, which are the equivalent of the Wilson coefficients in (\ref{MSbar}).
Matching conditions between
the effective field theories of (\ref{MSbar})
and $\Gamma_{\Delta S=1}^{LD}$ are obtained
by requiring 
that $S$-matrix elements
of asymptotic states are the same at some perturbative scale.
\be
\label{matching}
\langle 2 | \Gamma_{\Delta S=1}^{LD} | 1 \rangle =
\langle 2 | \Gamma_{\Delta S=1} | 1 \rangle \, .
\ee
The matching conditions 
fix {\em analytically} the short-distance behavior of the
couplings $g_i$
\ba
g_i(\mu_c,\cdots)&=& {\cal F}(C_i(\nu), \alpha_s(\nu), \cdots) \, .
\ea
This was done explicitly in  \cite{scheme} for
$\Delta S=2$ transitions and used in   \cite{epsprime}
for $\Delta S=1$ transitions.

\subsection{The Heavy $X$-Boson Method}

For energies below  the charm quark mass, we use
an effective field theory of heavy color-singlet
$X$-bosons coupled to QCD currents and
densities\cite{epsprime,scheme,DeltaI=1/2}.
For instance, the effective action reproducing
\begin{displaymath}
Q_1(x)=\left[\overline s \gamma^\mu (1-\gamma_5)  d \right] \,
 \left[\overline u \gamma_\mu (1-\gamma_5)  u \right](x)
\end{displaymath}
is 
\ba
\Gamma_{X}&\equiv& g_1(\mu_c,\cdots) 
\int {\rm d}^4 y 
\, X_1^\mu \left\{ \left[\overline s \gamma_\mu (1-\gamma_5)  
d \right](x) \right. \nonumber \\ &+& \left. 
 \left[\overline u \gamma_\mu (1-\gamma_5)  u \right](x) \right\}
\, .
\ea
Here the degrees of freedom
of quarks and gluons above the scale $\mu_c$ have been integrated out.
The advantage of this method is that two-quark currents are unambiguously
identified and that QCD densities are much easier to match
than four-quark operators. 

We use a 4-dimensional Euclidean cut-off $\mu_c$ 
to regulate UV divergences.
We can now calculate $\Delta S=1$ Green's functions
with the $X$-boson effective theory consistently.

\section{Long-Distance--Short-Distance Matching}

Let's study the $\Delta S=1$ two-point function
$$
\Pi(q^2)\equiv i\!\! \int {\rm d}^4 x \, e^{iq \cdot x}
\langle 0 | T(P_i^\dagger(0) P_j(x) e^{i \Gamma_{X}}|0\rangle .
$$
The $P_i$ are pseudoscalar sources
with quantum numbers describing $K\to  \pi$
amplitudes. 

Taylor expanding the off-shell amplitudes
$K\to\pi$ obtained from these Green's functions,
in external momentum and $\pi$, and $K$ masses,
one can obtain the couplings of the CHPT Lagrangian. 
These predict $K\to\pi\pi$
at a given order. This is unambiguous.

At leading order in the $1/N_c$ expansion the contribution  to the 
Green function $\Pi(q^2)$ is factorizable.
This only involves strong two-point functions
and is model independent.

The non-factorizable contribution, is NLO in the $1/N_c$
expansion. It involves the integration \cite{BK} of strong four-point functions
$\Pi_{P_iP_jJ_aJ_b}$ over the momentum Euclidean 
$r_E$
 that flows through the currents/densities $J_a$ and $J_b$
from 0 to $\infty$, schematically written as
\be
\Pi(q^2) \sim \int \frac{{\rm d}^4 r_E}{(2\pi)^4} 
\Pi_{P_iP_jJ_aJ_b} (q_E, r_E)  .
\ee
We separate long- from short-distance physics with
a cut-off $\mu$ in $r_E$. The short distance
part can be treated within OPE QCD. 

Recently, it was emphasized that 
dimension eight operators may be numerically important
for low values of the cut-off scale\cite{CDG00}. 
This issue can be studied straightforwardly in our approach.

There is {\em no} model dependence in our evaluation
of $K\to \pi$ amplitudes at NLO in $1/N_c$ within QCD
up to now.
The long distance part from 0 up to $\mu$ remains
For very small values of $\mu$ one can use CHPT
but it starts to be insufficient
already at relatively small values 
of $\mu$. Too small 
to match with the short-distance part.
The first step to enlarge the CHPT domain is to use a good hadronic model
for intermediate energies. We used the ENJL model\cite{ENJL}.
It has several good features
-it includes CHPT to order $p^4$, for instance-
and also some drawbacks as explained in \cite{scheme}.
Work is in progress to implement the 
large $N_c$ constraints on three-
and four-point functions along the lines of \cite{PPR98}.

\section{$\varepsilon'_K$ in the Chiral Limit}

\begin{figure}%1
\epsfxsize120pt
%\figurebox{120pt}{160pt}{}
\includegraphics[height=0.46\textwidth,angle=270]{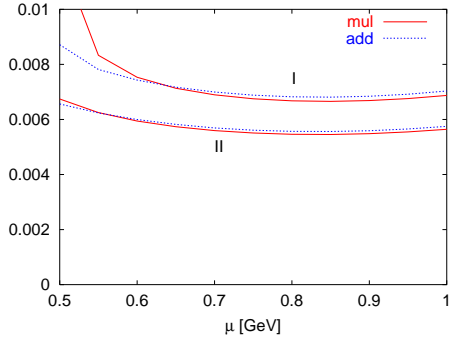}
\caption{Plot showing the matching between short- and long-distances
for $\varepsilon_K'/\varepsilon_K$.
Notice the wide plateau. Curves I and II correspond to different
choices in the running of $\alpha_s$ to two-loops.}
\label{fig:radk}
\end{figure}

To a very good approximation,
\ba
\left| \varepsilon_K'\right| &\simeq&
\frac{1}{ \, \sqrt 2} \frac{\re a_2}{\re a_0}\, 
\left\{ -\frac{\im a_0}{\re a_0}+\frac{\im a_2}{\re a_2}\right\}\, .
\ea
The lowest order CHPT values for $\re a_0$ and $\re a_2$ are
obtained from a fit\cite{KMW91}
to $K\to \pi\pi$ and $K\to \pi\pi\pi$
amplitudes to order $p^4$.
Our results\cite{DeltaI=1/2} reproduce 
the $\Delta I=1/2$ enhancement within  40 \%.
We use the {\it experimental} lowest order CHPT values\cite{KMW91} for 
$\re a_I$ to predict $\varepsilon_K'$ as shown in Figure \ref{fig:radk}.

For the two dominant operators and $\varepsilon_K'/\varepsilon_K$
we obtain
at NLO in  $1/N_c$ and in the chiral limit\cite{epsprime,DeltaI=1/2}
\ba
B_{6\chi}^{(1/2)NDR}(2 \, {\rm GeV})= 2.5 \, \pm 0.4 \nonumber \\
B_{8\chi}^{(3/2)NDR}(2 \, {\rm GeV})= 1.35 \, \pm 0.20 \nonumber\\
\label{chiral}
\left|\frac{\varepsilon_K'}{\varepsilon_K}\right|_\chi
= (60 \pm 30 )\cdot 10^{-4}\,.  
\ea

\section{Higher Order CHPT Corrections}

The r\^ole of FSI
in the standard\cite{lastresults} 
predictions
of 
$\varepsilon_K'/\varepsilon_K$ has been recently
studied\cite{FSI}. 
We took a different strategy.
The ratio  $\im a_I / \re a_I$ has no FSI
to all orders, thus FSI only affects the
ratio $\re a_2/\re a_0.$

Among the isospin breaking effects
only  $\pi^0$-$\eta$ mixing 
is under control and is known \cite{pi0eta}
 to order $p^4$. Other real $p^4$ and 
higher electromagnetic corrections are 
mostly~unknown.

Including $\pi^0$-$\eta$ mixing
and FSI our result (\ref{chiral})
becomes
\ba
\left|\frac{\varepsilon_K'}{\varepsilon_K}\right|
&=& (34 \pm 18 )\cdot 10^{-4} \, .
\ea

\section*{Acknowledgments}
This work is partially supported by the European Union TMR
Network EURODAPHNE (Contract No. ERBFMX-CT98-0169),
the Swedish Science Foundation (NFR), CICYT, Spain
(Grant No. AEN-96/1672), and Junta de Andaluc\'{\i}a
(Grant No. FQM-101).

\end{document}